\documentclass[preprint2]{aastex}

\input{psfig.sty}

\def\arcs{\rlap{.}$^{\prime\prime}$}
\def\arc{$^{\prime\prime}$}

\begin{document}

%% LaTeX will automatically break titles if they run longer than
%% one line. However, you may use \\ to force a line break if
%% you desire.

\title{A Technique for Separating the Gravitational Torques of
Bars and Spirals in Disk Galaxies}

%% Use \author, \affil, and the \and command to format
%% author and affiliation information.
%% Note that \email has replaced the old \authoremail command
%% from AASTeX v4.0. You can use \email to mark an email address
%% anywhere in the paper, not just in the front matter.
%% As in the title, you can use \\ to force line breaks.

\author{R. Buta\altaffilmark{1}, D. L. Block\altaffilmark{2}, and J. H. Knapen\altaffilmark{3}} 

%\email{aastex-help@aas.org}

%% Notice that each of these authors has alternate affiliations, which
%% are identified by the \altaffilmark after each name.  Specify alternate
%% affiliation information with \altaffiltext, with one command per each
%% affiliation.

\altaffiltext{1}{Department of Physics and Astronomy, University of Alabama, Box 870324, Tuscaloosa, AL 35487 USA}
\altaffiltext{2}{School of Computational and Applied Mathematics, University of the Witwatersrand, P.O. Box 60, Wits, Gauteng 2050, South Africa}
\altaffiltext{3}{Department of Physical Sciences, University of Hertfordshire,
Hatfield, Herts AL10 9AB, UK}

%% Mark off your abstract in the ``abstract'' environment. In the manuscript
%% style, abstract will output a Received/Accepted line after the
%% title and affiliation information. No date will appear since the author
%% does not have this information. The dates will be filled in by the
%% editorial office after submission.

\begin{abstract}
We describe a Fourier-based method of separating bars from spirals
in near-infrared images. The method
takes advantage of the fact that a bar is typically a feature with
a relatively fixed position angle, and uses the simple assumption
that the relative Fourier amplitudes due to the bar decline with radius
past a maximum in the same or a similar manner as they rose to that 
maximum. With
such an assumption, the bar can be extrapolated into the spiral
region and removed from an image, leaving just the spiral and
the axisymmetric background disk light. We refer to such a bar-subtracted
image as the "spiral plus disk" image. The axisymmetric background (Fourier
index $m$=0 image) can then be added back to the bar image to
give the "bar plus disk" image. The procedure allows us to estimate
the maximum gravitational torque per unit mass per unit square of the
circular speed for the bar and spiral forcing separately, parameters which
quantitatively define the bar strength $Q_b$ and the spiral strength $Q_s$
following the recent study of Buta \& Block.
For the first time, we are able to measure the torques
generated by spiral arms alone, and we can now define spiral
torque classes, in the same manner as bar torque classes
are delineated.

We outline the complete procedure here using a 2.1$\mu$m image of NGC 6951, 
a prototypical SAB(rs)bc spiral having an absolute blue magnitude of $-$21 and
a maximum rotation velocity of 230 km s$^{-1}$. Comparison between
a rotation curve predicted from the $m$=0 near-infrared light distribution
and an observed rotation curve suggests that NGC 6951 is maximum disk
in its bar and main spiral region, implying that our assumption of a 
constant mass-to-light ratio in our analysis is probably reliable. 
We justify our assumption on how to make the bar extrapolation using an
analysis of NGC 4394, a barred spiral with only weak near-infrared spiral 
structure, and we justify the number of needed Fourier terms using NGC 1530, 
one of the most strongly-barred galaxies (bar class 7) known. We also
evaluate the main uncertainties in the technique. Allowing for uncertainties
in vertical scaleheight, bar extrapolation, sky subtraction, orientation
parameters, and the asymmetry in the spiral arms themselves, we estimate
$Q_b$=0.28$\pm$0.04 and $Q_s$=0.21$\pm$0.06 for NGC 6951.

\end{abstract}

%% Keywords should appear after the \end{abstract} command. The uncommented
%% example has been keyed in ApJ style. See the instructions to authors
%% for the journal to which you are submitting your paper to determine
%% what keyword punctuation is appropriate.


%% From the front matter, we move on to the body of the paper.
%% In the first two sections, notice the use of the natbib \citep
%% and \citet commands to identify citations.  The citations are
%% tied to the reference list via symbolic KEYs. The KEY corresponds
%% to the KEY in the \bibitem in the reference list below. We have
%% chosen the first three characters of the first author's name plus
%% the last two numeral of the year of publication as our KEY for
%% each reference.

\section{Introduction}

In two previous papers (Buta \& Block 2001, hereafter BB01; Block et
al. 2001), we outlined a new approach to quantifying the observed bar
strengths of galaxies. Instead of relying on deprojected bar
ellipticities, we used a theoretical equation (Combes \& Sanders 1981)
based on the forcing of the bar implied by the near-infrared light
distribution. This method, called the relative bar torque, or $Q_b$,
method, was applied to 36 galaxies by BB01 and later to nearly 40 more
galaxies by Block et al. (2001). The method has also been applied to
more than 100 Two Micron All Sky Survey (2MASS) galaxies by
Laurikainen, Salo, \& Rautiainen (2002) and by Laurikainen \& Salo
(2002), who also refined the method. Most recently, Block et
al. (2002) applied the $Q_b$ method to more than 150 galaxies in the
Ohio State University Bright Galaxy Survey (OSUBGS, Eskridge et al.
2002) and used the results to show that normal galaxies may double
their mass by accretion in 10$^{10}$ years.

A difficulty with the $Q_b$ method as applied in these previous
studies is that the bar strengths based on the method could be
affected by spiral arm torques. $Q_b$
represents the maximum ratio of the tangential force to the
mean axisymmetric radial force. If the bar is a typical SB-type bar
(rather than an oval),
then this maximum will usually be a good approximation to the
bar strength and the force ratio map will show a characteristic butterfly
pattern (BB01). However, many barred spirals have pronounced spiral
arms that break directly from the ends of the bar, and these arms
can affect the bar butterfly pattern and increase the apparent
bar strength. 

The original intent of BB01 was that $Q_b$ should measure bar strength,
not a combination of bar and spiral arm strength. There are good
reasons for trying to find a way to separate the effects of the
spiral from the bar. First, if we want to investigate scenarios
of bar formation in disk galaxies (e.g., Sellwood 2000), 
then we should have measures of
bar torques, not spiral plus bar torques. Second, with a separation
analysis, we can check theoretical predictions that bars with larger
torques drive spirals with higher amplitudes (Elmegreen \& Elmegreen
1985 and references therein; also see Yuan \& Kuo 1997 and references
therein). 

In this paper, we outline a straightforward method of separating
the bars from the spirals using 2.1$\mu$m near-infrared images. 
The method uses Fourier
techniques in conjunction with a simple assumption of how to
extrapolate the bar into the spiral-dominated regions. The method
works effectively even for the strongest bars with the strongest
spirals. For the first time, we can investigate spiral arm
torques in galaxies and even define spiral torque classes in the
same manner as BB01 defined bar torque classes.

We illustrate the method using a representative example, the SAB(rs)bc
spiral NGC 6951. The image we use is a $K$-short, or $K_s$, image
obtained during a run with the 4.2-m William Herschel Telescope (WHT) in 2001.
The image scale is 0\arcs 24 per pixel and the field of view is
4\rlap{.}$^{\prime}$1 square.
A total of 15 spiral galaxies was observed for investigating bar and
spiral torques. Full details of these observations, and analysis of the
remaining galaxies, will be provided in Block et al. (2003, hereafter
paper II) where the techniques described in this paper will be applied
to examine the relations between bars and spirals.

\section{Estimation of Gravitational Torques}

The $Q_b$ method was fully described in BB01. The dimensionless
parameter $Q_b$ can be interpreted as the maximum gravitational bar
torque per unit mass per unit square of the circular speed.  To derive
it, we process a near-infrared image by removing all foreground stars,
and then deproject the image using available orientation
parameters. For NGC 6951, we used a mean position angle $<\phi>$ and
axis ratio $<q>$ based on isophotal ellipse fits on the 2.1$\mu$m
image itself. The values used were $<q>$ = 0.773 and $<\phi>$ =
143\rlap{.}$^{\circ}$1.  These are in good agreement with optical
photometric estimates of the same parameters from M\'arquez \& Moles
(1993).  IRAF routine IMLINTRAN was used for the deprojection. To
facilitate our analysis, we have rotated the deprojected image such
that the bar axis is horizontal. The bar position angle in the raw
deprojected image was measured using ellipse fits, and IRAF routine
ROTATE was used for the final rotation.

The deprojected image was then centered within an array of dimension
2$^n$, where $n$=10 for the WHT image, and run through a program which
transforms the near-infrared image into a two-dimensional potential
(Quillen, Frogel, \& Gonz\'alez 1994, hereafter QFG). From the 2D
potential, planar forces are calculated and then decomposed into
radial and tangential components.  Our main analysis is based on maps
of the ratio of the tangential force to the mean axisymmetric radial
force, the latter derived from the $m$=0 component of the
potential. We assume a constant mass-to-light ratio, but in addition
to the force ratios, we also compute a predicted axisymmetric rotation
curve in order to evaluate the correctness of this assumption,
especially in the spiral arm regions. We have also made a refinement
to our use of the QFG potential method, based on a study by
Laurikainen \& Salo (2002), who noted that the QFG convolution
integral for the vertical dimension included some gravity
softening. We use a revised lookup table from H. Salo (private
communication) for an exponential vertical density profile without
softening. Laurikainen \& Salo (2002) showed that the relative bar
torques of BB01 are too low by about one bar class because of this
softening.

The computation of a potential from a near-infrared image requires a
value for the vertical scaleheight, which cannot be directly measured
for NGC 6951. In BB01 and Block et al. (2001), it was assumed that all
galaxies had the same vertical exponential scaleheight as our Galaxy,
$h_z$ = 325pc. However, this approach required knowledge of the
distance to each galaxy, which had to be based on radial
velocities. Here we derive $h_z$ by scaling a value from the radial
scalelength, $h_R$. As shown by de Grijs (1998), the ratio $h_R$/$h_z$
(based on mostly $I$-band and some $K$-band surface photometry)
depends on Hubble type, being larger for later types compared to
earlier types. For NGC 6951, we estimated a radial exponential
scalelength using an azimuthally-averaged $K_s$ surface brightness
profile. The slope of the outer light profile provides an
approximation to a radial scalelength, which we obtained to be $h_R$ =
33\arc. From a bulge/disk decomposition of an $I$-band luminosity
profile, M\'arquez \& Moles (1993) obtained a disk effective radius of
42\rlap{.}$^{\prime\prime}$76 for NGC 6951, which corresponds to a
radial exponential scalelength of 25\rlap{.}$^{\prime\prime}$5. This
is in reasonable agreement with our estimate.  For an Sbc spiral, de
Grijs's analysis shows that $h_R$ = (6$\pm$2)$h_z$ on average. For a
redshift distance of 24.1 Mpc (Tully 1988) for NGC 6951, this gives
$h_z$ = 640pc, about twice the Galactic value. The average value of
$h_z$ in de Grijs's sample is 600$\pm$400pc.

We found it necessary to remove some of the very strong star-forming
regions from the near-infrared image of NGC 6951 before the potential
was calculated.  These objects can cause local maxima or minima in the
force ratio maps that may be unreliable if the mass-to-light ratios of
these regions differ from the dominant old stellar background. It is
well-known that HII regions and luminous red supergiants impact the
2.2$\mu$m spectral region (e.g., Knapen et al. 1995), and can be
locally important at the 33\% level (Rhoads 1998).

\section{Bar/Spiral Separation: A Fourier Approach}

The basic idea of our approach is that the bar is a feature dominated 
mostly by even Fourier terms in a relatively fixed position angle.
We compute the relative Fourier intensity amplitudes $I_m/I_0$, 
where $m$ is an integer index, as a function of radius and make the 
assumption, when necessary, that the relative bar intensity declines 
past a maximum in the same or a similar manner as it rises
to that maximum. That is, the Fourier amplitudes relative to the
axisymmetric background have a single maximum at radius $r_m$ and
decline smoothly and roughly symmetrically to either side of this radius.
The bar extrapolation involves scaling the
sine and cosine amplitudes of the even Fourier terms according to the
ratio, $I_0(r_m+\Delta r)/I_0(r_m-\Delta r)$,
of the $m$=0 amplitudes at symmetric radii $r_m\pm\Delta r$.
This is essential because the average intensity of the background
starlight decreases with increasing radius and because we extrapolate 
$I_m/I_0$, not $I_m$ alone. As long as the Fourier phases are relatively 
constant, this scaling is reasonable.

Our assumption about how to make the extrapolations is based on
data from mostly pure bar plus disk systems, and on cases where the
bar and spiral are well-enough separated that we can differentiate their 
distinct contributions in plots of $I_m/I_0$ and the $m$=2 phase $\phi_2$. 
Ohta, Hamabe, \& Wakamatsu (1990) presented
relative Fourier amplitudes for six early-type barred systems that
showed three characteristics of the relative Fourier amplitudes of
bars: (1) strong bars have significant higher order terms, such that even 
$m$=10
can be important; (2) relative amplitudes rise and then decline past
a maximum that lies roughly in the middle of the apparent bar; (3) 
the radii of the maxima for higher order terms ($m$ = 4, 6, 8, etc.)
can shift to slightly larger values compared to $m$ = 2.
Correct treatment requires that we examine
plots of the relative amplitudes of all even Fourier terms used, to
evaluate especially effects (2) and (3).

\begin{figure*}
%\vspace{10cm}
\psfig{figure={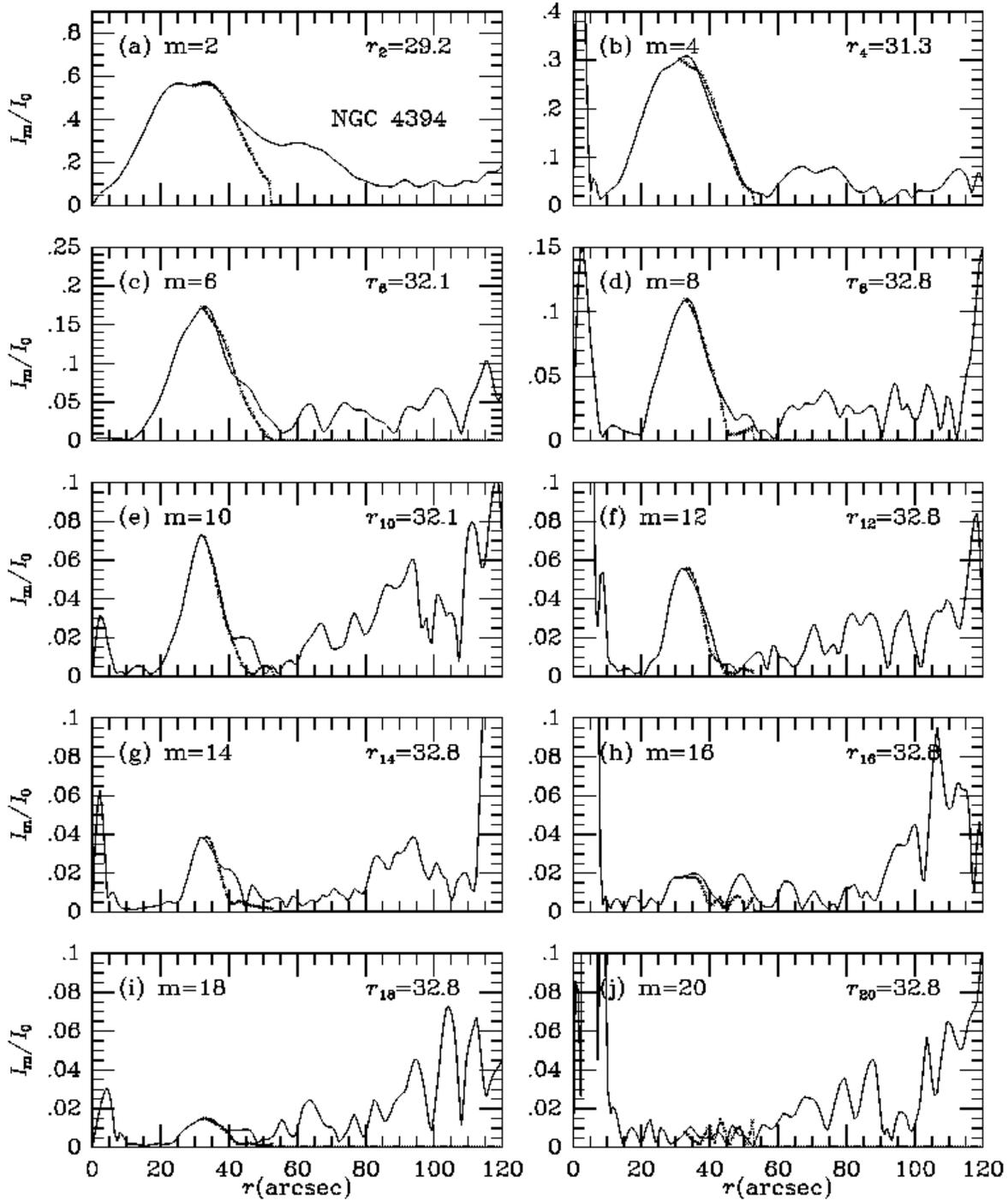},width=.95\textwidth}
\vspace{0.5cm}
\caption{Relative Fourier intensity amplitudes as a function of
radius for NGC 4394, for even terms to $m$=20. Solid curves
show the observed relative amplitudes for each term, while the
small crosses show how well reflecting the rising amplitudes
for $r$ $<$ $r_m$ match the observed declines for $r$ $>$
$r_m$, where $r_m$ (in arcseconds) is indicated in each panel.
The different behavior of the $m$=2 term is discussed in the text.}
\label{n4394}
\end{figure*}

To illustrate these points, we use the galaxy NGC 4394 from the Ohio
State University Bright Galaxy Survey (OSUBGS, Eskridge et al. 2002).
Even in blue light, the spiral structure of this galaxy is weak,
and in the OSUBGS $H$-band image, 
it is quite subdued. Thus, NGC 4394
can serve as our model case of mostly a bar imbedded in a disk, and we
can examine how the relative Fourier amplitudes behave with radius.
Figure~\ref{n4394} shows the even relative Fourier amplitudes to $m$=20
for NGC 4394 versus radius based on a deprojected version of the OSUBGS 
$H$-band image where the bulge has been properly treated with a 
two-dimensional 
decomposition (Laurikainen et al. 2003).
Each Fourier term shows a well-defined maximum in the bar region. 
When the relative Fourier intensities are extrapolated past this maximum
in the same manner as they rose to that maximum, the extrapolations are
seen to be in fairly good agreement with the observed relative intensities.
There is weak spiral structure near and outside the ends of the bar, but 
it contributes little to the higher order terms. Detailed examination
of the radius, $r_m$, of the peak in each plot shows that it increases to
slightly larger values with increasing $m$, as expected from point 3 above.
The idea, according to Ohta, Hamabe, \& Wakamatsu (1990), is that
the narrow ends of the bar occur at the largest radii, and will be
most evident as a result in the higher-order terms at those larger
radii. Finally, the phases of the terms (not shown) are relatively constant 
throughout the bar. Thus, NGC 4394 demonstrates that our assumption of 
approximate symmetry for the even Fourier terms as a function of 
radius for a bar is a fair one.

The symmetry assumption is less evidently correct for the $m$=2
term in NGC 4394. Instead, this term shows a strong asymmetric decline
past its maximum. The reason for this difference is that the main bar in NGC 
4394 is imbedded in a weak extended oval that has no higher order terms
than $m$=2. There is also some weak spiral structure at the
ends of the bar. For these reasons, the $m$=2 term in NGC 4394
would have to be extrapolated as shown in Figure~\ref{n4394} 
to isolate the main bar. 

The $m$=10 term in NGC 4394 has a maximum about 13\% of the maximum
of the $m$=2 term. Thus, it is not really negligible compared
to $m$=2, verifying point 2 above. For a very strong bar, such as
that in NGC 1530, the $m$=10 term is 19\% of the 
maximum relative amplitude of the $m$=2 term. One should never
think naively of a bar as an $m$=2 structure. Only a broad
oval is likely to be a pure $m$=2 structure. 

Once the Fourier mappings of the bar are determined, the next step
in the procedure is to sum the even ($m\geq$2) terms of the light 
distribution out to a specified radius, where we assume the bar goes 
to zero; there is also usually an inner radius where the bar is
assumed to go to zero. Both of these effects are seen clearly for
NGC 4394 in Figure~\ref{n4394}, where terms $m$=4 and 6 approach 
zero near $r$=10$^{\prime\prime}$ and $r$=55$^{\prime\prime}$. Terms
$m$=8, 10, and 12 approach zero near $r$=20$^{\prime\prime}$.
For most of the even terms in NGC 4394, there appears to be some amplitude
inside 10$^{\prime\prime}$ due in part in the finite pixel size
(1\rlap{.}$^{\prime\prime}$5). Each Fourier term is treated on an 
individual basis to allow the radius of the maximum, $r_m$, to
occur at a different position. 

The maximum number of Fourier terms we use is based on analysis of
NGC 1530, the most strongly barred galaxy in the WHT sample.
Using the BB01 approach, NGC 1530 would have been bar class 7.
To deduce how many Fourier terms might be needed for a
bar/spiral separation analysis in this kind of case, we analyzed
NGC 1530 by comparing gravitational torques derived from a
full resolution image with those derived from Fourier-smoothed
images which cut the terms at $m$=10 and $m$=20. We found that
except for the effects of noise, the Fourier-smoothed image 
with terms to $m$=20 adequately represented the bar and that
including higher order terms is not essential. Cutting the analysis 
at $m$=10 was not sufficient in that case. However, for many
galaxies, only the lower-order terms would be needed for
a separation analysis, and we allow for the option of cutting
the number of terms when appropriate.

\section{Illustration of Technique}

\begin{figure*}
%\vspace{10cm}
\psfig{figure={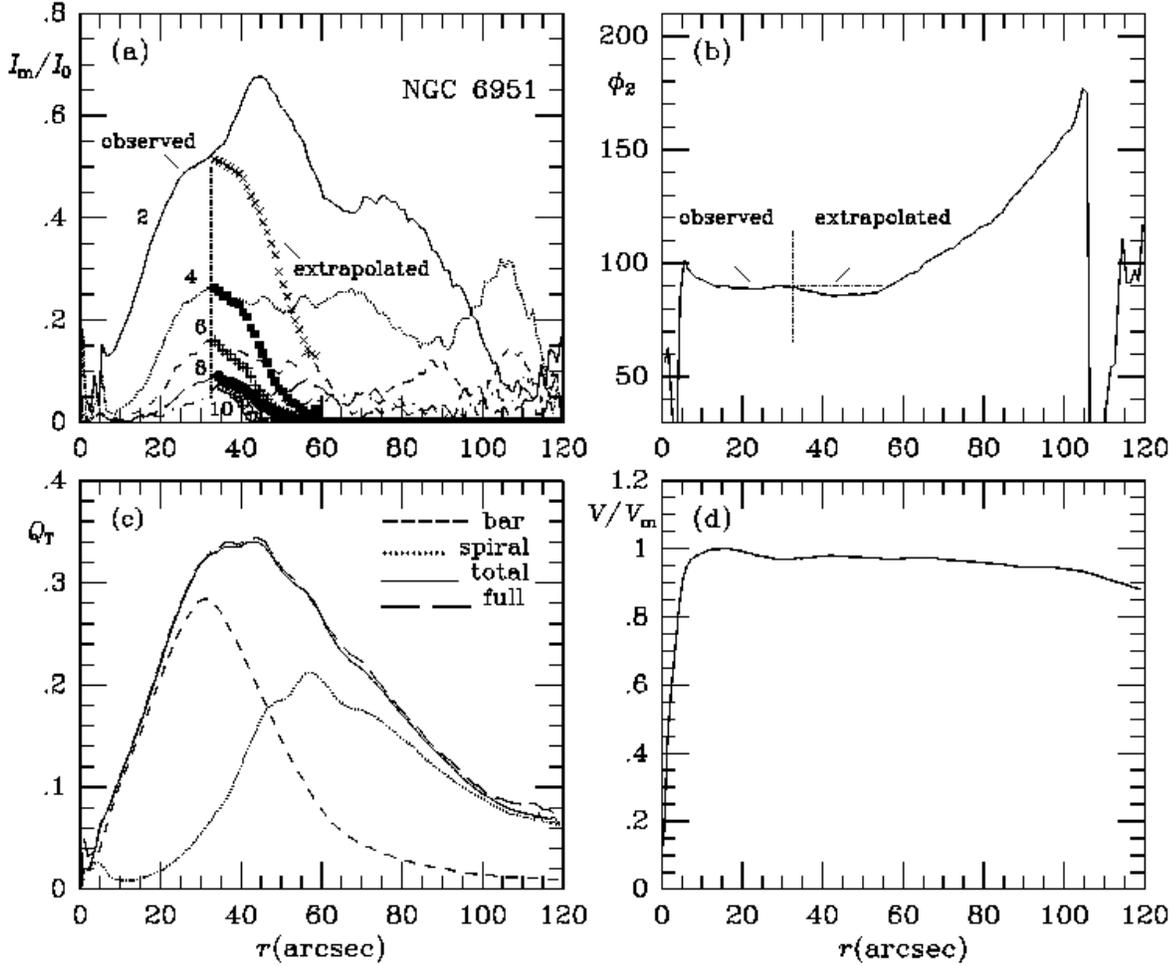},width=.95\textwidth}
\vspace{0.5cm}
\caption{Analysis plots  for NGC 6951: (a)  relative Fourier intensity
amplitudes  for $m$=2, 4,  6, 8,  and 10  as a  function of  radius in
arcseconds.  The vertical dotted line (radius $r_2$=32\arcs 5) divides
the  observed from  the extrapolated  $m$=2 relative  amplitudes.  The
symbols show  the extrapolations for  each Fourier term, and  the plot
highlights the  slight shift in the  radius of the maximum  in the bar
region as  $m$ increases (at  least for $m$=8  and 10). (b)  Phase (in
degrees)  of the  $m$=2  component as  a  function of  radius. In  the
bar-dominated  region  from  $\approx$10\arc\  to  32\arc,  the  phase
$\phi_2$  is  relatively  constant.  The extrapolation  reflects  this
constancy  around the radius,  $r_2$, of  the input  maximum (vertical
dotted line).   (c) Mean  maximum relative gravitational  torque $Q_T$
versus  radius for  the  bar (dashed  curve),  spiral (dotted  curve),
$m$=0-20 sum image (solid  curve), and full image (long-dashed curve).
(d) predicted  rotation curve (normalized  to the maximum,  $V_m$) for
NGC  6951 from  the  $K_s$ image,  assuming  a constant  mass-to-light
ratio.}
\label{plot6951}
\end{figure*}

\begin{figure*}
%\vspace{10cm}
\psfig{figure={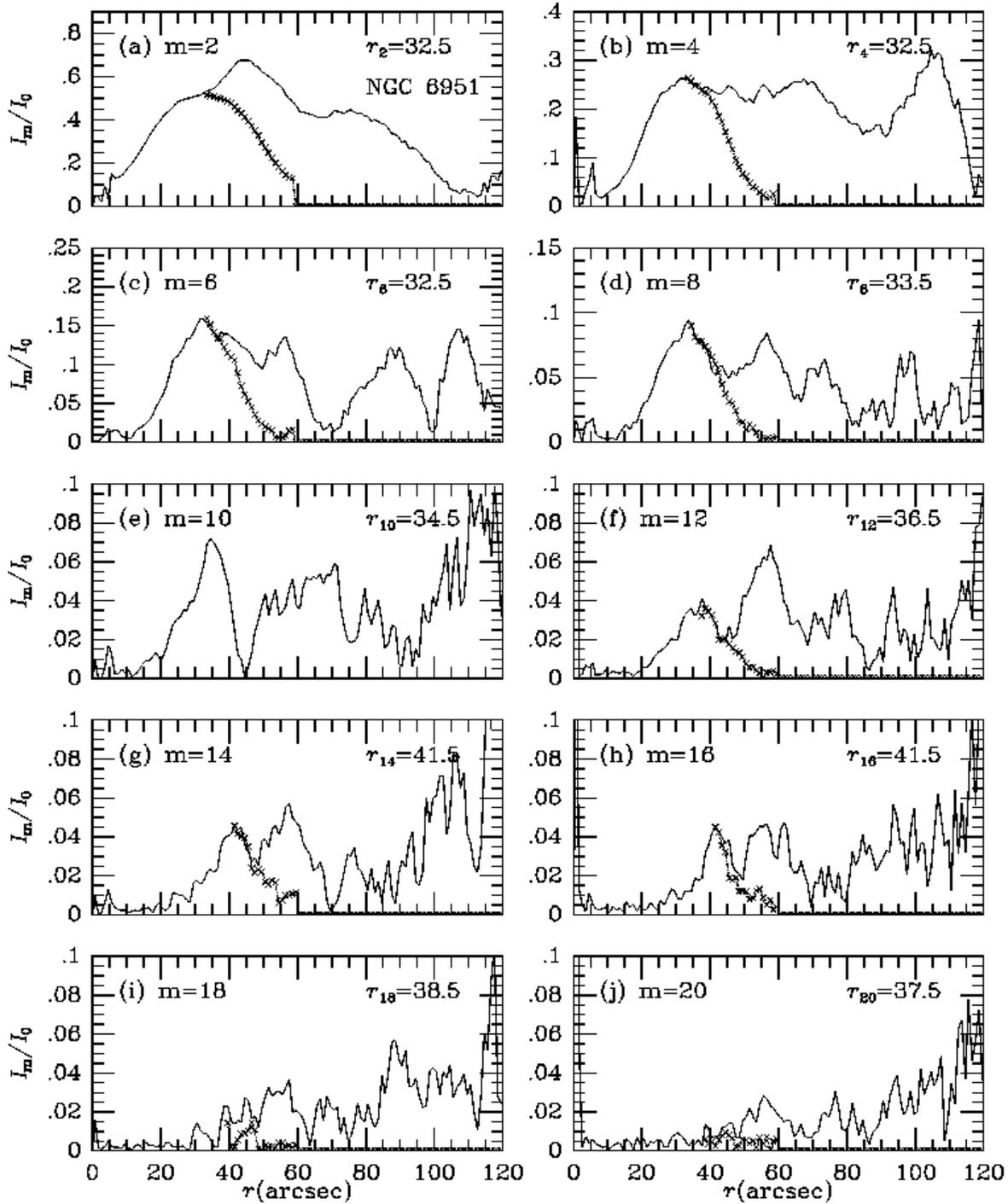},width=.95\textwidth}
\vspace{0.5cm}
\caption{Relative Fourier intensity amplitudes as a function of radius
for  NGC 6951, for  even terms  to $m$=20.  Extrapolations of  the bar
terms  are  shown by  the  small crosses,  and  the  radius $r_m$  (in
arcseconds) is indicated in each panel. For $m$=10, the amplitudes are
used as observed  for the decline past $r_m$,  and are extrapolated to
zero for $r$ $>$ 45$^{\prime\prime}$ (see text).}
\label{n6951mplots}
\end{figure*}

\begin{figure*}
\centerline{\psfig{figure={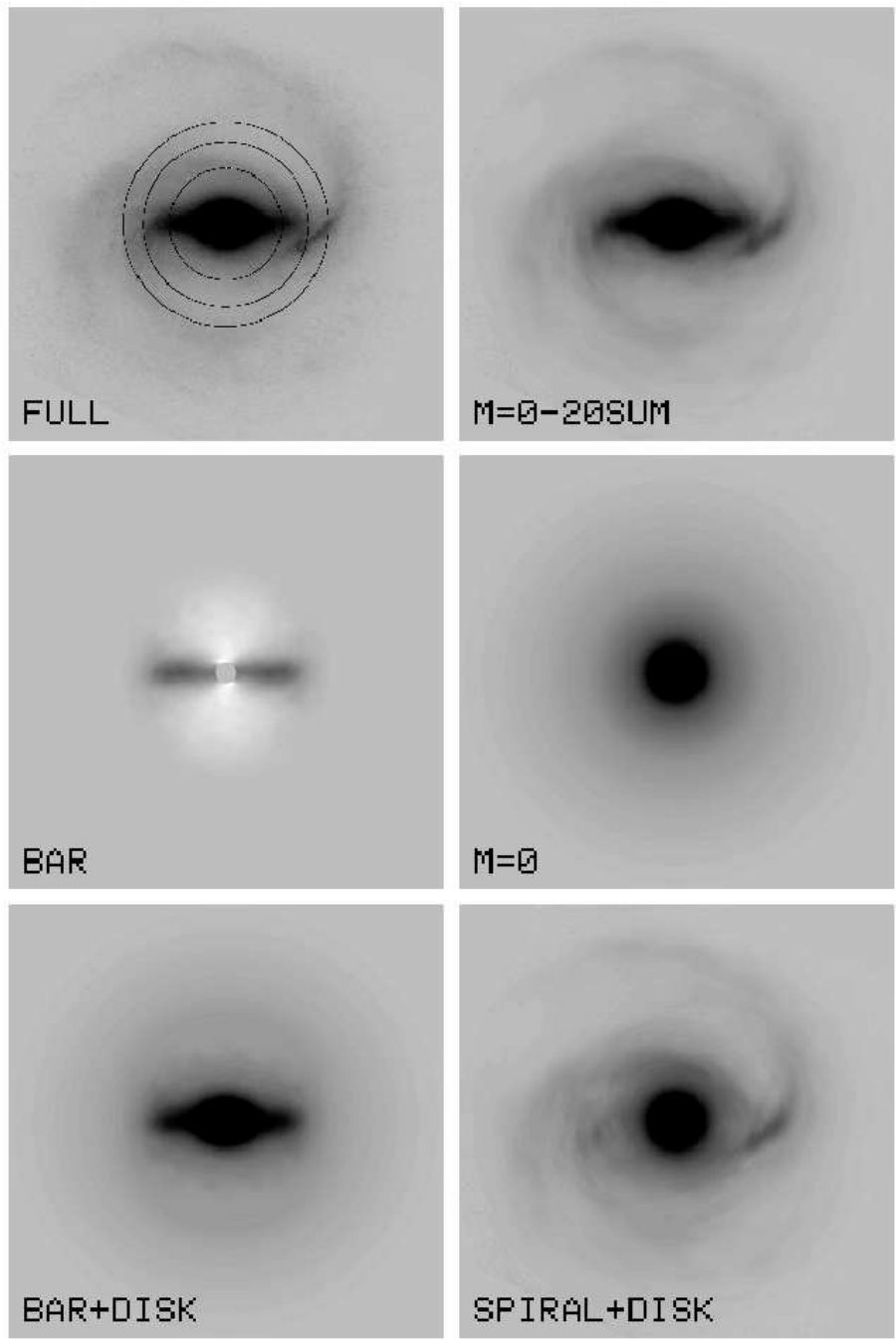},width=.73\textwidth}}
\vspace{0.5cm}
\caption{Various images used for the bar/spiral separation
analysis of NGC 6951. Each frame indicates the type of
image shown. The full image is the original deprojected, 
rotated image. The $m$=0-20 sum is the Fourier-smoothed image
based on 21 Fourier terms (including $m$=0 and both even and
odd terms). The bar image 
is based on the extrapolations shown in Figure 2 and includes
all even terms from $m$=2 to $m$=20. The $m$=0 image is
the mean axisymmetric background. The bar+disk and spiral+disk
images show the separated components against the mean axisymmetric
background. The three circles superposed on the full image correspond,
in increasing radius, to $r(Q_b)$, the $Q_T(bar)$=$Q_T(spiral)$ 
crossover radius, and $r(Q_s)$.}
\label{n6951images}
\end{figure*}

\begin{figure*}
%\vspace{10cm}
\centerline{\psfig{figure={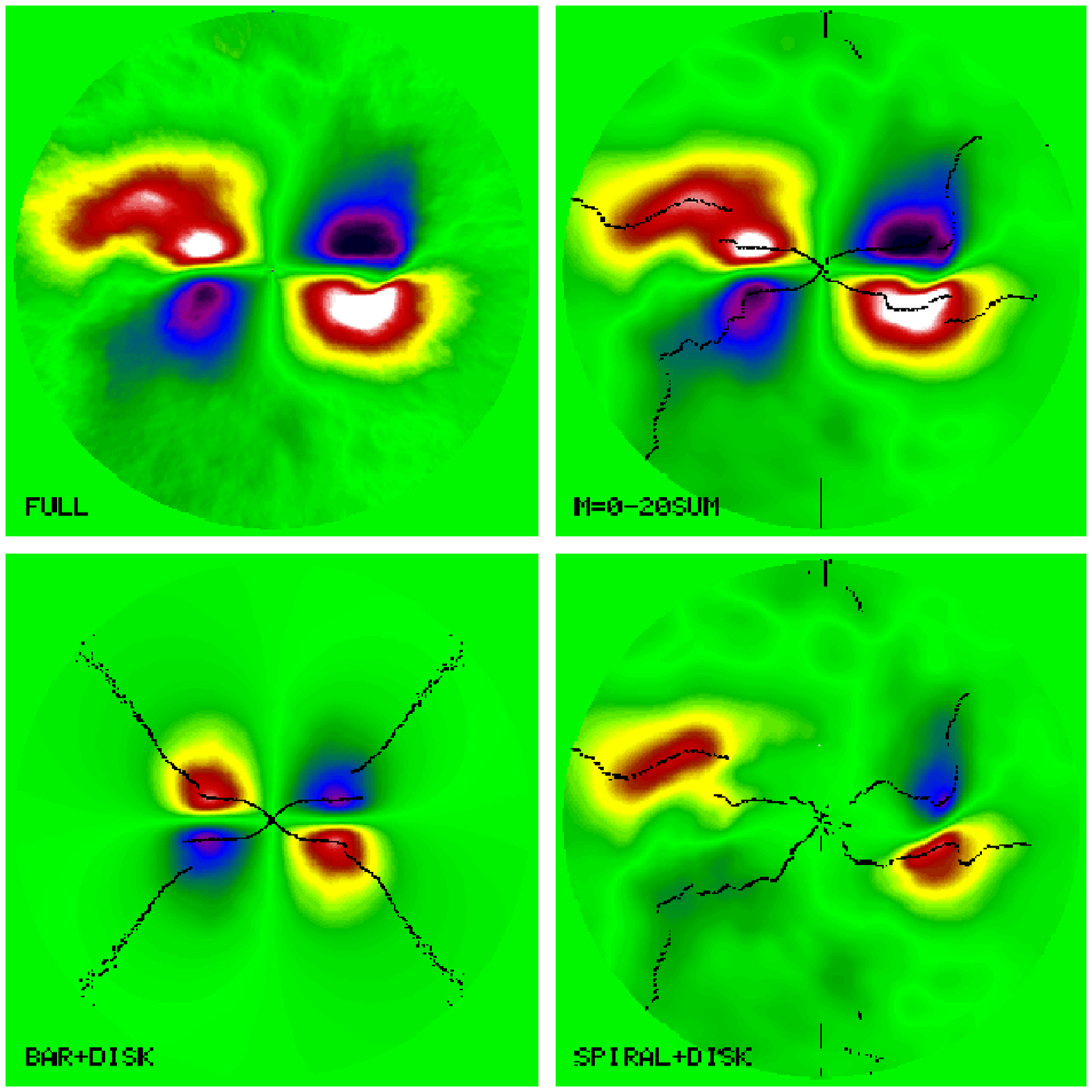},width=\textwidth}}
\vspace{0.5cm}
\caption{Color-coded ratio maps of the tangential force to the
mean axisymmetric radial force for four of the images of NGC 6951
in Figure 3. These maps are equivalent to the gravitational torque
per unit mass per unit square of the circular speed.
The reddish-white zones are areas where this force ratio is positive
while the bluish-purple zones are areas where this force ratio is
negative. The dark curves show mappings of $|Q_T^{max}|$ in each
quadrant relative to the bar or the spiral.}
\label{n6951ratios}
\end{figure*}

The method is illustrated for NGC 6951 in Figures~\ref{plot6951},
~\ref{n6951mplots},
~\ref{n6951images}, and ~\ref{n6951ratios}. NGC 6951 is an ideal
test case because it has a well-defined bar, and its spiral structure
breaks directly from the ends of this bar. It is a very typical example.
In Figure~\ref{plot6951}a,
we show the relative $K_s$ Fourier intensity amplitues for the $m$=2, 4, 6, 8, 
and 10 components. (Figure~\ref{n6951mplots} shows the relative amplitudes
to $m$=20 on a larger scale, as for NGC 4394.) 
Between radii of 6\arc\ and 32\arc, the bar 
dominates these amplitudes and we see a smooth rise in most of the terms. 
Beyond 32\arc, the relative $m$=2 amplitude rises to a higher maximum, 
and the $m$=2 phase decreases slightly and then rises (see 
Figure~\ref{plot6951}b). We assume the bar does not end abruptly,
but that in the absence of the spiral, the even relative Fourier
amplitudes due to the bar decline in the same manner as they rose 
to the bar maximum, which we took to occur at $r_2$ = 
32\rlap{.}$^{\prime\prime}$5. This was chosen because $I_2/I_0$
shows a plateau near this radius, and the higher-order terms all
show a maximum near or just outside this radius as well. It
corresponds to a position approximately in the apparent middle 
of the bar. We require that the bar still be
significant in $m$ = 2 at its apparent edge, and allow it drop in relative
$m$=2 amplitude sharply thereafter. This extrapolates the bar into the spiral 
zone (as indicated in Figure~\ref{plot6951}a). The impact of the choice of
$r_2$ is examined in the next section.

For the higher order terms, Figure~\ref{n6951mplots} shows clear maxima
between 33\arc\ and 42\arc\ that are attributable mainly to the bar.
These maxima do appear to shift outwards a little with increasing $m$
(especially for $m>$ 6), as expected from point (3) above. For these terms, 
we simply reflect the even amplitudes around the apparent radius $r_m$
where the maxima occur, and
scale them according to the mean intensity at each radius outside
$r_m$. For NGC 6951, these extrapolations are shown by the crosses
in Figure~\ref{n6951mplots}.

Figure~\ref{n6951mplots} also shows that Fourier terms to $m$=18 are
still detectable in the bar of NGC 6951. Including such terms provides
a very good approximation to the total galaxy image.
To verify this, we computed potentials
from both the full image at maximum resolution, and a Fourier-smoothed
image based on the sum of all even and odd terms up to $m$=20.  
We found that the Fourier-smoothed $m$=0-20 image
gives virtually the same potential as the full image, with differences
being mainly attributable to noise and the occasional bright star-forming
region.

Interestingly, Figure~\ref{n6951mplots} shows that the $m$=10 term goes to zero
at $r$=45\arc, while our extrapolations make the other terms go to
zero at 55\arc\ to 60\arc. This shows the limitations of our approach
in the sense that the bar relative intensity profiles may not be as
symmetric as we assume. None of the terms in NGC 4394 shows a similar
disagreement with the other terms, so $m$=10 may be unusual in 
NGC 6951. Our procedure in general is that if any
term shows a significant portion of its decline past the maximum, we
will use the decline as measured and extrapolate as little as possible.
This is what is done for $m$=10 in Figure~\ref{n6951mplots}.
However, we have tested that our maximum relative torque results for NGC 6951 would
be the same even if the $m$=10 term were extrapolated as for the other
terms. The impact of cutting all the even-order $m>2$ terms at 
$r$=45$^{\prime\prime}$ is considered in the next section.

With extrapolations defined for each Fourier term, we computed the 
images shown in Figure~\ref{n6951images}. The ``full" image is the original
deprojected and rotated image, with the bar horizontal. This is the
full resolution image and includes all noise. The ``$m$=0-20 sum" is the
Fourier-smoothed version of the full image. It is based on 21 Fourier
terms, including all even and odd terms. The ``bar" image is the one
that uses the extrapolations shown in Figure~\ref{n6951mplots}. It was
computed by summing the $m\geq$2 Fourier terms within the limits set for
the bar, and has no net flux.\footnote{All $m$$>$0 terms in a Fourier
series, when integrated over all azimuths, have no net flux. The net
flux in the bar would be the sum of all of its higher-order terms plus
its $m$=0 term. The net flux in the bar is brought back when we add
the $m$=0 image to the ``bar'' image.} The fourth image is the $m$=0 Fourier
image, which shows the axisymmetric part of the $K_s$ light distribution.

The remaining images in Figure~\ref{n6951images} show the separated bar
and spiral images. The image ``bar+disk" is the sum of the ``bar" and
$m$=0 images. The image ``spiral+disk" is the ``$m$=0-20 sum" minus the
``bar" image. With this procedure, we place most of the
noise and all of the odd Fourier terms into the ``spiral+disk" image. 

Figure~\ref{n6951ratios} shows the force ratio maps 
$Q(i,j)=F_T(i,j)/F_{0R}(i,j)$, where $F_T$ is the tangential force
and $F_{0R}$ is the mean axisymmetric radial force,
for the full, Fourier-smoothed, bar+disk, and spiral+disk images of
NGC 6951. In the full image, one clearly sees the effects of both
the spiral and the bar. A dominant ``butterfly" pattern is present
with extra structure due to the spiral. The Fourier-smoothed
image looks very much the same, only with less noise. In the bar+disk
ratio map, we see four symmetrically placed ``maximum points"
which lie near the ends of the bar, as noted in BB01. The spiral+disk
image shows a less symmetric, rough butterfly pattern as well.
Thus, our approach appears to have separated the bar and spiral,
and we can see how each component contributes to the total ratio
map.

Our next step is to use the ratio maps to derive single numbers
that characterize the bar strength and the spiral strength. For this
purpose, we derive the maximum value, $Q_T$, of the ratio of the
tangential force to the mean radial force as a function of radius.
These curves are interpolated bilinearly from the force ratio maps.
Figure~\ref{n6951ratios} shows that the structure in the force
ratio maps alternates by quadrants with two maximum positive
values and two maximum negative values at each radius. BB01
derived $|Q_T(max)|$ in each quadrant, and then averaged these four
values to get a single number, $Q_b$, called the ``bar strength"
in the galaxy. Here we use a similar procedure, but as a function
of radius. In Figure~\ref{n6951ratios}, the dotted curves show the
maxima(minima) for each quadrant. For the bar+disk, a symmetric
pattern for these maxima(minima) is mapped, as expected, while
for the spiral a more complex mapping is found. Because the spiral
is more complicated than the bar, we divided the ratio map into
two 180$^{\circ}$ sections around a vertical line, and searched for
the maxima and minima in each section. The sharp breaks in the
spiral mapping are discontinuities in azimuth but not in radius,
and are attributable to structure in the arms. At each radius,
we averaged the values of $|Q_T(max)|$ to get the plots shown
in Figure~\ref{plot6951}c. The dashed curve shows that the bar
has a maximum force ratio at $r(Q_b)$=31\arcs 5 (3.7 kpc), similar to
the value chosen from the relative intensity amplitudes (smallest
circle superposed on the full image in Figure~\ref{n6951images}).
The spiral maximum in the dotted curve occurs at $r(Q_s)$=57\arcs 5
(6.7 kpc, largest circle superposed on the full image).
For comparison, the solid curve in Figure~\ref{plot6951}c
shows the maxima from the total Fourier-smoothed image, while
the crosses show the maxima from the full image. These
curves agree well and show again that 21 Fourier terms
are adequate for the analysis of NGC 6951. 

Figure~\ref{plot6951}c shows that $Q_T$(bar)=$Q_T$(spiral) at
$r$$\approx$46\arcs 5. This corresponds to the middle circle
superposed on the full image in Figure~\ref{n6951images}.
The circle lies just outside the ends of the bar, and passes
through the bright spiral arc on the right side of the bar.

From these curves we derive $Q_b$=0.284$\pm$0.001 and $Q_s$=0.212$\pm$0.035,
where the uncertainties include only the internal scatter in maximum values.
The small internal uncertainty in $Q_b$ is due to the fact that 
only even Fourier terms were used to define the bar. The larger internal
uncertainty in $Q_s$ is due to the fact that all Fourier terms, even and odd,
were used for the spiral. Additional uncertainties are discussed
in the next section.

The maximum total relative gravitational torque is $Q_g$=0.340. Thus, our analysis
shows that in this case, ignoring the effects of the spiral
would lead us to overestimate the bar strength by about
20\%. Nevertheless, the BB01 ``bar class" remains at 3 
in either case.

Figure~\ref{plot6951}d
shows the predicted, normalized rotation curve for NGC 6951, based on the
$m$=0 term of the derived gravitational potential. The apparent
flatness of the predicted curve for radii beyond $r$=10$^{\prime\prime}$
is consistent with observed rotation
profiles along the photometric major axis obtained by M\'arquez \& Moles (1993)
and P\'erez et al.
(2000), and supports our assumption of a constant mass-to-light
ratio, at least for this galaxy. The maximum rotation velocity
in NGC 6951 is $V_m$=230 km s$^{-1}$ (M\'arquez \& Moles 1993), 
favoring a maximal disk according to Kranz, Slyz, \& Rix (2003).

\section{Uncertainties in Method}

BB01 discussed in detail the uncertainties in estimating
gravitational torques from near-infrared images. One of the principal
uncertainties is in the vertical scaleheight $h_z$. Since this parameter
can be measured directly only for edge-on galaxies,
it has to be assumed. As shown by de Grijs (1998), $h_R/h_z (=2h_R/z_0$,
where $z_0$ is the isothermal scaleheight) depends
on Hubble type, but within any type there is a significant scatter.
At stage Sbc, individual values of this ratio range from 4 to 8.
This range must include both a cosmic scatter component as well as
fitting and observational uncertainties. In our estimated value of 
$Q_b$ for NGC 6951, the scatter in the ratio leads to an
uncertainty of $\pm$0.022, while for $Q_s$ it leads to an uncertainty
of $\pm$0.020. Thus, the uncertainty due to vertical scaleheight
is at the 8-9\% level for both $Q_b$ and $Q_s$. This excludes the
uncertainty in our derived value of $h_R$ for NGC 6951, which is
sensitive to tke sky subtraction. This is discussed further below,
but note again that some of the uncertainty in this parameter due to 
observational errors and decomposition uncertainties must be included 
in the spread in $h_R/h_z$ at each type.

Related to the issue of scaleheights is the possibility that 
the bar in NGC 6951 might be thicker than its disk, an effect
thought to cause a boxy/peanut-shaped bulge structure in some
edge-on disk galaxies (Bureau \& Freeman 1999). However, Laurikainen \& Salo 
(2002) showed that such a non-constant vertical scaleheight is
unimportant in the evaluation of relative bar torques, finding 
that it affects $Q_b$ by less than 5\%. Since we cannot be sure
what the vertical structure of the bar in NGC 6951 is like 
compared to its disk, we assume this could contribute
an uncertainty of 0.05$Q_b$ for our estimate of $Q_b$.

\begin{figure*}
%\vspace{10cm}
\psfig{figure={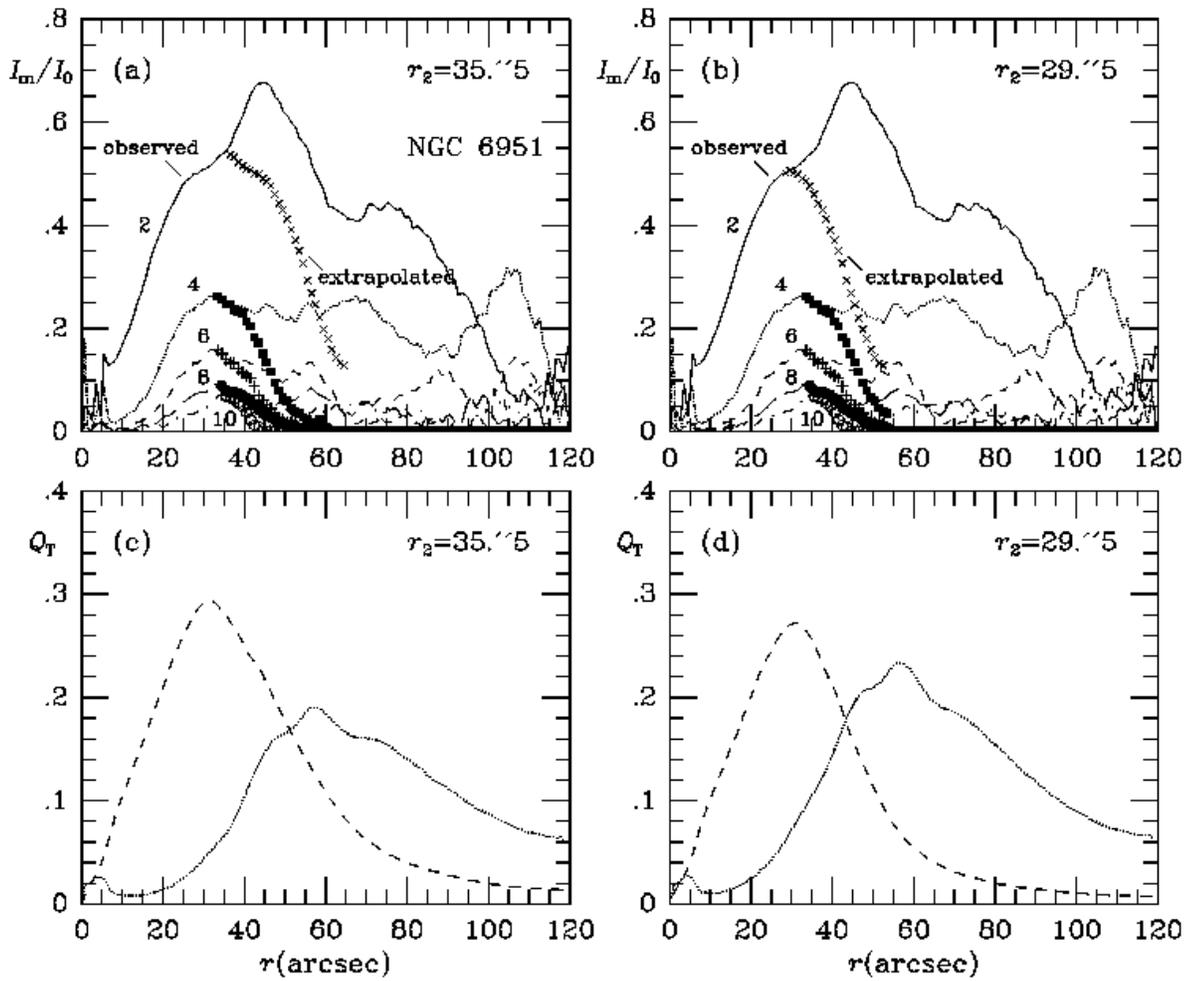},width=.95\textwidth}
\vspace{0.5cm}
\caption{Analysis plots of NGC 6951 for different extrapolations of the
$m$=2 component: (a), (c) for $r_2$ = 35\arcs 5; (b), (d) for $r_2$ = 
29\arcs 5. In (c) and (d), the dashed curve is for the bar and the
dotted curve is for the spiral. See Figure 1 caption for further
explanations.}
\label{unc}
\end{figure*}

For bar/spiral separation, another possible major uncertainty will be the
extrapolations of the bar Fourier amplitudes. The largest Fourier term 
in any bar will be the $m$=2 term. We first investigate how the choice of the 
radius of the $m$=2 maximum impacts our results. For our analysis of NGC 6951
in the previous section, we used 32\arcs 5 for this radius, and obtained,
as already noted, $Q_b$=0.284$\pm$0.001 and $Q_s$=0.212$\pm$0.035. If
we use 35\arcs 5 instead (Figure~\ref{unc}a and c), 
we obtain $Q_b$=0.293 and $Q_s$=0.190, while
if we use 29\arcs 5 (Figure~\ref{unc}b and d), 
we obtain $Q_b$=0.272 and $Q_s$=0.234. Thus, the uncertainties
in $Q_b$ and $Q_s$ due to the extrapolation will be correlated
in the sense that if $Q_b$ is too high, then $Q_s$ will be too low
and vice-versa. The choice of the radius of the $m$=2 maximum appears to affect
$Q_s$ more than $Q_b$ for NGC 6951. A $\pm$10\% uncertainty in the
radius of the bar maximum leads to $\pm$4\% uncertainty in $Q_b$
and $\pm$10\% uncertainty in $Q_s$. Note that the greatly
different $m$=2 extrapolations have little or no impact on the
radii of the bar and spiral maxima. Both extrapolations still
give $r(Q_b)$$\approx$31\arc\ and $r(Q_s)$$\approx$57\arc. However,
the radius of the crossover point, where $Q_T(bar)=Q_T(spiral)$,
is sensitive to the $m$=2 extrapolation, ranging from
52\arc\ in Figure~\ref{unc}c to 43\arc\ in Figure~\ref{unc}d.

The second issue connected with the bar extrapolation is the symmetry
assumption of the relative amplitudes. We have noted that 
the $m$=10 term in NGC 6951 seems to violate this assumption and goes
to zero at $r$=45$^{\prime\prime}$, while our assumed extrapolations 
for the other terms go to zero at larger radii. To test the impact 
of the symmetry assumption, we use the same extrapolations as before
for all even
$m$$\neq$10, but cut all $m>$2 at $r$=45$^{\prime\prime}$. This approximates
the asymmetry of the $m$=10 term for all even $m>$2. 
Cutting the $m$=2 term at the same radius appears to be too drastic,
however, because it leaves a sharp edge in the separated bar and spiral images.
Thus, we have left the extrapolation for $m$=2 the same as before for this
test. The cutoff for the higher-order terms
changes the derived relative maximum torques to $Q_b$ = 
0.281 and $Q_s$ = 0.220, amounting to 1\% for $Q_b$ and 4\% for $Q_s$.
This shows that, even if the terms actually do cut off at a radius smaller
than implied by our extrapolations, we do not commit 
a serious error in $Q_b$ and $Q_s$ if we ignore it.

Other uncertainties discussed by BB01 included the flattening of the
bulge, bulge deprojection stretch, uncertainties in the orientation
parameters, sky subtraction uncertainties, and
the constant $M/L$ assumption. NGC 6951 has a low enough inclination
that bulge deprojection stretch has a negligible effect on the derived
maximum torques, which occur for both the bar and spiral far outside the
bulge-dominated area. Uncertainties due to
orientation parameters can be significant (see Table 3 of BB01).
Buta et al. (2003) show that, for a galaxy inclined about 40$^{\circ}$,
the typical uncertainties in relative maximum torques is $\pm$0.030
for $\pm$5$^{\circ}$ uncertainty in inclination and $\pm$4$^{\circ}$ 
uncertainty in major axis position angle.

The favorable agreement between the observed
and predicted rotation curves of NGC 6951 suggests that dark matter
is not important in the inner parts of the galaxy, and that
our assumption of a constant $M/L$ is probably correct in this
case. In general, the best way to evaluate the effects of dark matter 
on relative maximum torque calculations will be to compare predicted 
near-infrared rotation curves with observed ones, if available.

Sky subtraction errors could impact our torque calculations, 
because the field of view of near-infrared images is usually
limited and the sky level cannot always be precisely determined.
For NGC 6951, we estimate a sky level  uncertainty of $\pm$0.2 ADU
in the intensity scale of the $K_s$ image. Such an uncertainty in 
the sky level will naturally affect our estimate of $h_R$ and hence 
also $h_z$. We derive an uncertainty of $\pm$4$^{\prime\prime}$
in $h_R$ for $\pm$0.2 ADU uncertainty in the sky, corresponding to an 
uncertainty in the vertical scaleheight of $\pm$78pc if $h_z$=(1/6)$h_R$.
This combined sky subtraction/vertical scaleheight uncertainty
leads to an uncertainty of $\pm$0.010 in $Q_b$ and $\pm$0.011 in $Q_s$.
As might have been expected, the sky subtraction error impacts
$Q_s$ more than $Q_b$ since the arms lie at larger radii, although
the effect is only slight. The change is 3.5\% for $Q_b$ and 5\% for $Q_s$. 

Thus, allowing for the uncertainties in vertical scaleheight, bar
extrapolation, sky subtraction, orientation parameter uncertainties,
and the asymmetry in the spiral arms themselves,
we derive $Q_b$=0.284$\pm$0.040 and $Q_s$=0.212$\pm$0.056 for
NGC 6951.

Finally, there will always be individual cases that require special
treatment, particularly if the bar has considerable asymmetry. In
such cases, allowance for odd Fourier terms in the bar is needed, and
we will describe in paper II an approach to dealing with them.

\section{Conclusions}

We have shown that straightforward Fourier techniques can be used
to make a reasonable separation of a bar from a spiral. With such
a separation, we can extend the BB01 gravitational bar torque method
to spirals and define a spiral strength as well as a bar strength.
In the case of NGC 6951, the maximum relative total gravitational torque
is 0.34, while the individual bar and spiral strengths are 0.28$\pm$0.04
and 0.21$\pm$0.06, respectively. Thus, NGC 6951 is bar class 3 and spiral class
2, following Table 1 of BB01.

The general applicability of our separation method has not been
fully evaluated since it has only been used for our sample of
15 WHT galaxies and two additional galaxies from other sources.
However, as we will show in paper II, a reasonable separation was
possible in each of those cases. We anticipate that the 
method will be applicable to most spirals, but that
some galaxies will require special treatment. Objects having multiple
bars, very weak bars in strong ovals, or very asymmetric bars are
examples of such cases. We consider some of these cases in paper II.

We rename the maximum relative total gravitational torque parameter as $Q_g$,
to remove any ambiguity about what it represents. 
In cases with strong bars and weaker
spirals, the actual bar strength $Q_b$ $\approx$ $Q_g$, while in
cases with strong spirals and weaker bars, $Q_s$ $\approx$ $Q_g$.
In general, a separation analysis will be needed to investigate
real bar torques. However, the derivation of $Q_g$ alone is useful
for some studies (e.g., investigations of the impact of gas accretion
in galaxy disks; Block et al. 2003), and is the most straightforward
parameter to derive. More details on the practical aspects of
deriving $Q_g$ for a large number of galaxies is provided by
Laurikainen et al. (2003). 

We thank H. Salo and E. Laurikainen for helpful discussions concerning
refinements of the $Q_b$ method, and for providing a revised
lookup table for the vertical dimension in our analysis and for the
deprojected OSUBGS $H$-band image of NGC 4394.
We also thank an anonymous referee for several helpful comments.
RB acknowledges the support of NSF grant AST-0205143 to the University
of Alabama and also the Anglo American Chairman's fund during
a visit to the University of the Witwatersrand in 2002 November.
The research of DLB is supported by the Anglo American Chairman's Fund,
with deep gratitude expressed to CEO M. Keeton, Special Advisor
H. Rix, and the Board of Trustees. DLB is also most appreciative
to SASOL for funding the sabbatical period of his research. The William 
Herschel Telescope is operated on the island of La Palma by the Isaac Newton 
Group in the Spanish Observatorio del Roque de los Muchachos of the 
Instituto de Astrof\'\i sica de Canarias. Funding for the OSU Bright
Galaxy Survey was provided by grants from the National Science
Foundation (grants AST-9217716 and AST-9617006), with additional
funding from the Ohio State University.

\clearpage

\centerline{REFERENCES}

\noindent
Bureau, M. \& Freeman, K. C. 1999, \aj, 118, 126

\noindent
Buta, R. \& Block, D. L. 2001, \apj, 550, 243 (BB01)

\noindent
Block, D. L., Puerari, I., Knapen, J. H., Elmegreen, B. G., Buta, R.,
Stedman, S., \& Elmegreen, D. M. 2001, \aap, 375, 761

\noindent
Block, D. L., Bournaud, F., Combes, F., Puerari, I., \& Buta, R. 2002,
\aap, 394, L35

\noindent
Block, D. L., Buta, R., Knapen, J. H., Elmegreen, D. M., Elmegreen, B. G.,
Puerari, I., \& Stedman, S. 2003, in preparation

\noindent
Combes, F. \& Sanders, R. H. 1981, \aap, 96, 164

\noindent
de Grijs, R. 1998, \mnras, 299, 595

\noindent
Elmegreen, B. G. \& Elmegreen, D. M. 1985, \apj, 288, 438

\noindent
Eskridge, P., Frogel, J. A., Pogge, R. W., et al. 2002, \apjs, 143, 73

Knapen, J. H., Beckman, J. E., Heller, C. H., Shlosman, I., \& de Jong, R. S. 
1995, \apj, 454, 623

\noindent
Kranz, T., Slyz, A., \& Rix, H.-W. 2003, \apj, 586, 143

\noindent
Laurikainen, E. \& Salo, H. 2002, \mnras, 337, 1118

\noindent
Laurikainen, E., Salo, H., \& Rautiainen, P. 2002, \mnras, 331, 880

\noindent
Laurikainen, E., Salo, H., Buta, R., \& Vasylyev, S. 2003, in preparation

\noindent
M\'arquez, I. \& Moles, M. 1993, \aj, 105, 2090

\noindent
Ohta, K., Hamabe, M., \& Wakamatsu, K. 1990, \apj, 357, 71

\noindent
P\'erez, E., M\'arquez, I., Marrero, I., Durret, F., Gonz\'alez Delgado, R. M.,
Masegosa, J., Maza, J., \& Moles, M. 2000, \aap, 353, 893

\noindent
Quillen, A. C., Frogel, J. A., \& Gonz\'alez, R. A. 1994, \apj, 437, 162 (QFG)

\noindent
Rhoads, J. E. 1998, \aj, 115, 472

\noindent
Sellwood, J. A. 2000, in Dynamics of Galaxies: From the Early Universe
to the Present, F. Combes, G. A. Mamon, \& V. Charmandaris, eds.,
San Francisco, ASP Conf. Ser. 197, p. 3.

\noindent
Tully, R. B. 1988, Nearby Galaxies Catalogue, Cambridge, Cambridge
University Press

\noindent
Yuan, C. \& Kuo, C.-L. 1997, \apj, 486, 750

\end{document}